\documentclass[aps, prb, reprint]{revtex4-1}
\usepackage{graphicx}

\begin{document}

\title{Near-field interaction of twisted split-ring resonators}

\author{David A. Powell}
\email{david.a.powell@anu.edu.au}
\author{Kirsty Hannam}
\author{Ilya V. Shadrivov}
\author{Yuri S. Kivshar}

\affiliation{Nonlinear Physics Centre, Research School of Physics and Engineering, Australian National University, Canberra, ACT 0200, Australia}

\begin{abstract}
We present experimental, numerical and analytical results for the study of near-field interaction of twisted split-ring resonators, the basic elements of the so-called \emph{stereometamaterials}. In contrast to previous results, we observe a crossing point in the dispersion curves where the symmetric and antisymmetric modes become degenerate. We introduce a model to describe the interplay between magnetic and electric near-field interactions and demonstrate how this model describes the crossing of the dispersion curves, initially considering lossless identical resonators.  Finally, we apply the theory of Morse critical points to demonstrate the competition between losses and fabrication errors in determining whether or not symmetric and antisymmetric modes cross.
\end{abstract}

\maketitle

\section{Introduction}

Metamaterials created as an array of sub-wavelength, resonant elements can exhibit interesting electromagnetic properties, such as a negative refractive index~\cite{Smith2000}.  An important \emph{building block} of metamaterials is the Split Ring Resonator (SRR)~\cite{Pendry1999}, which has a highly dispersive magnetic polarizability which is negative over some frequency band.  However the response of a metamaterial is not simply given by the response of an individual resonator, but depends also on the arrangement of resonators within the system\cite{Gay-Balmaz2002}.  Unlike the atoms in natural materials, the near-field patterns of metamaterial elements are quite complex, giving rise to strong interactions between them.  This means that the arrangement of resonators in a lattice or super-lattice plays an important role in determining the response of the metamaterial\cite{Gorkunov2002,Lapine2009,Powell2010,Feth2010,Sersic2009,Liu2008,Liu2008a}. A review of many examples of interaction between such resonant elements can be found in Ref.~\onlinecite{Liu2009a}. Characterizing the underlying interaction mechanisms between elements is essential to understanding the overall resonant properties and effective parameters of the material.  By controlling the relative arrangement of elements, it is possible to alter the response of the material substantially even if the constituents are fixed.

Recently, H.~Liu et al\cite{Liu2007} analyzed numerically an array consisting of pairs of split ring resonators on the same axis, with the second ring rotated by $90^\circ$.  The hybridized modes of this system show strong polarization rotation, and have suppressed radiation losses\cite{Li2009a}.  Subsequently, N.~Liu et al\cite{Liu2009} investigated a similar system, but considered an arbitrary angle between the two rings. Using numerical analysis, they extracted the eigenmodes of a pair of SRRs operating in the near infrared.  They concluded that as twist angle increases, the resonances converge, undergo an avoided crossing, then diverge again.  The reason for this avoided crossing is not clear, since our previous study of a different systems of coupled SRRs showed that the hybridized modes can cross\cite{Powell2010}.  In Ref.~\onlinecite{Liu2009} the dispersion curve with an avoided crossing was fitted by a multipole interaction model, however no physical justification for this fitting was given.  In particular, it was assumed that the magnetic interaction constant is invariant with twist angle, however this is inconsistent with the strong variation of current density around the circumference of the ring.  This model was subsequently extended to include the polarization-rotation of the scattered radiation\cite{Liu2010}.

The purpose of this article is twofold. First, we give a rigorous analysis of the interaction between a pair of co-axial SRRs with one ring rotated, and determine the conditions under which the dispersion curves will undergo a crossing or avoided crossing.  We introduce a physically-based model for describing the interaction between the rings, and show how this model explains the experimentally and numerically observed dispersion behavior. In Section~\ref{sec:experimental_numerical}, we present our experimental and numerical analysis of the dispersion curves.  In Section~\ref{sec:ideal_theory} we show how the dispersion curves can cross, based on an idealized model of a pair of SRRs.  Finally, in Section~\ref{sec:morse_theory} we consider the influence of losses and non-identical rings on the dispersion curves.  Using the theory of Morse critical points, we will show that competition between losses and fabrication imperfection will determine whether or not the modes cross.

\section{Experimental Results} \label{sec:experimental_numerical}

\begin{figure}[b]
	\centering
		\includegraphics[width=\columnwidth]{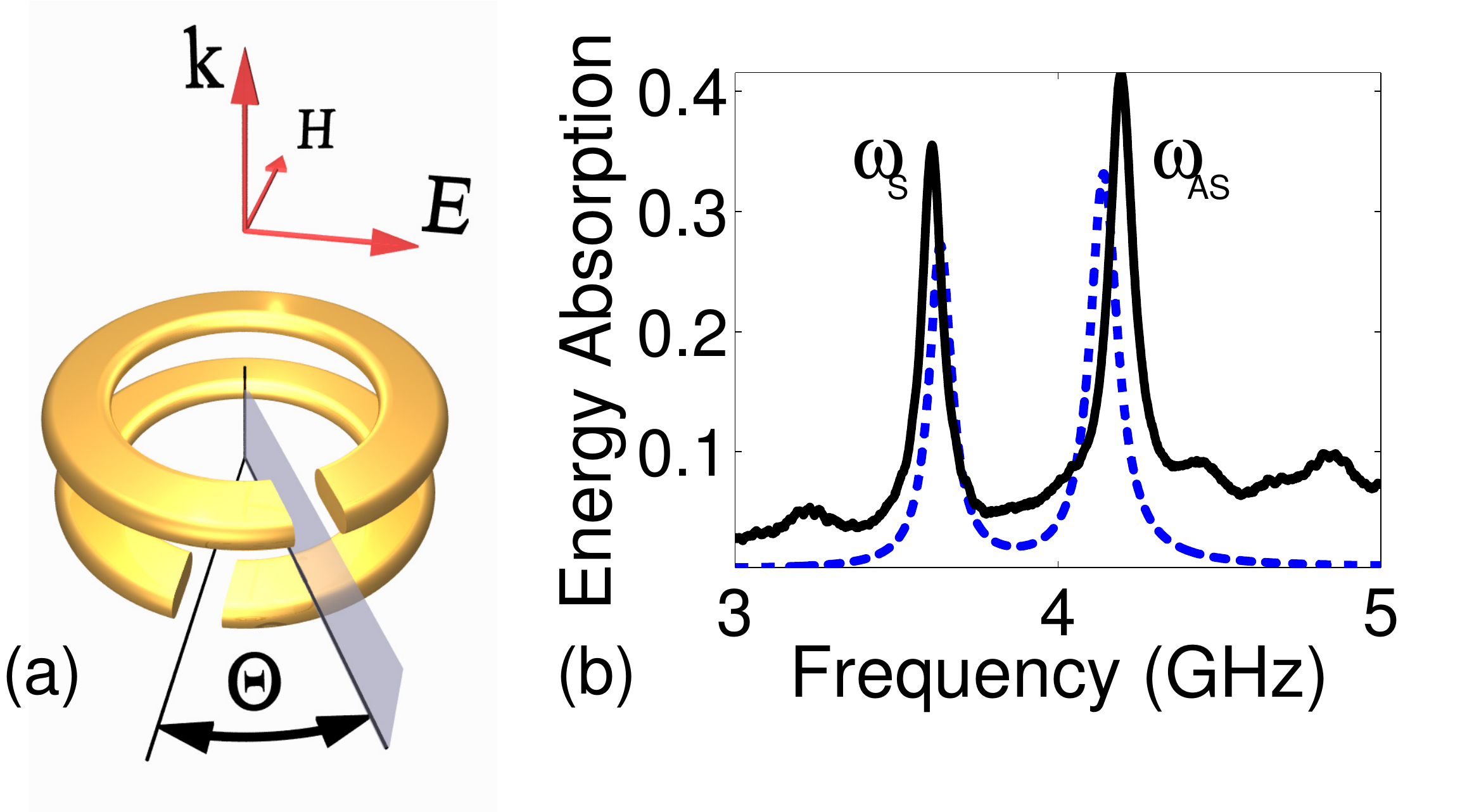}
	\caption{(a) A schematic showing the rings rotated with respect to each other through angle $\theta$, and the polarization of the incoming waves.  (b) A comparison of the experimental (solid) and numerical (dashed) absorption for angle $\theta = 90^\circ$.}
	\label{fig:Schematic}
\end{figure}

We consider a pair of SRRs with varying twist angle $\theta$ between them shown schematically in Fig.~\ref{fig:Schematic}(a). First, we perform microwave experiments with a series of rings, with one ring held fixed, and a separate sample created for each rotated ring.  The rings used have an inner radius of 3.5mm, an outer radius of 4mm, and a gap of 1mm.  They are copper, printed onto 1.6mm thick FR4 circuit board, and the rings 3.6mm apart, with the dielectric boards located between the rings.  The incoming microwaves are polarized so that the electric field is across the gap of the fixed ring, as shown in Fig.~\ref{fig:Schematic}(a).  To match our experimental data, we preform numerical calculations using CST Microwave Studio with the boards having a dielectric constant of 4.6.

\begin{figure}[b]
	\centering
		\includegraphics[width=\columnwidth]{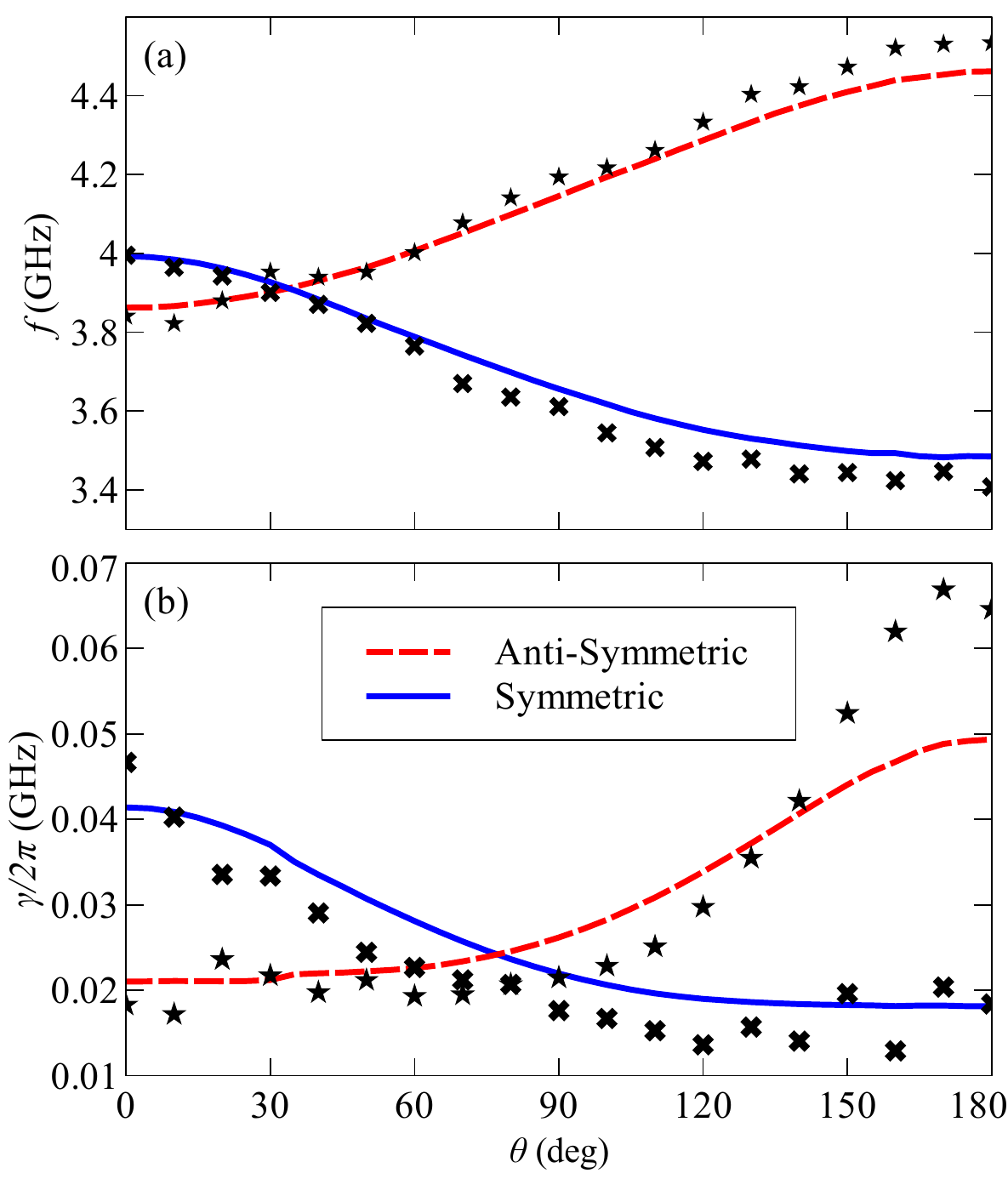}
	\caption{(a) A comparison of the experimental (markers) and numerical (lines) resonant frequencies. (b) Corresponding absorption coefficients, calculated from the resonance line-widths.}
	\label{fig:experimental_dispersion}
\end{figure}

Experimental results were measured using a Rohde and Schwarz ZVB network analyzer in a WR-229 rectangular waveguide. We measure the excitation of the rings from the absorption of the system, given by $1 - |S_{21}|^{2} - |S_{11}|^{2}$, where $S_{21}$ is the transmission coefficient, and $S_{11}$ is the reflection coefficient.  A comparison of the experimental and numerical absorption curves for $\theta = 90^\circ$ is shown in Fig.~\ref{fig:Schematic}(b). To characterize the dispersion behavior, experiments were performed with $\theta$ varied from $0^\circ$ to $180^\circ$ in $10^\circ$ increments, while numerical results were calculated in $5^\circ$ increments.  For each angle a Fano function\cite{Miroshnichenko2010} was fitted to the experimentally and numerically obtained absorption curves, and the resulting resonant frequencies and absorption coefficients are shown in Figs.~\ref{fig:experimental_dispersion}(a) and (b) respectively.  In all cases the data were well-described by a pair of Fano resonances with a correction for background absorption, and the fitting to the numerical results has a residual error below $5\times10^{-3}$.

\begin{figure}[bt]
	\centering
		\includegraphics[width=\columnwidth]{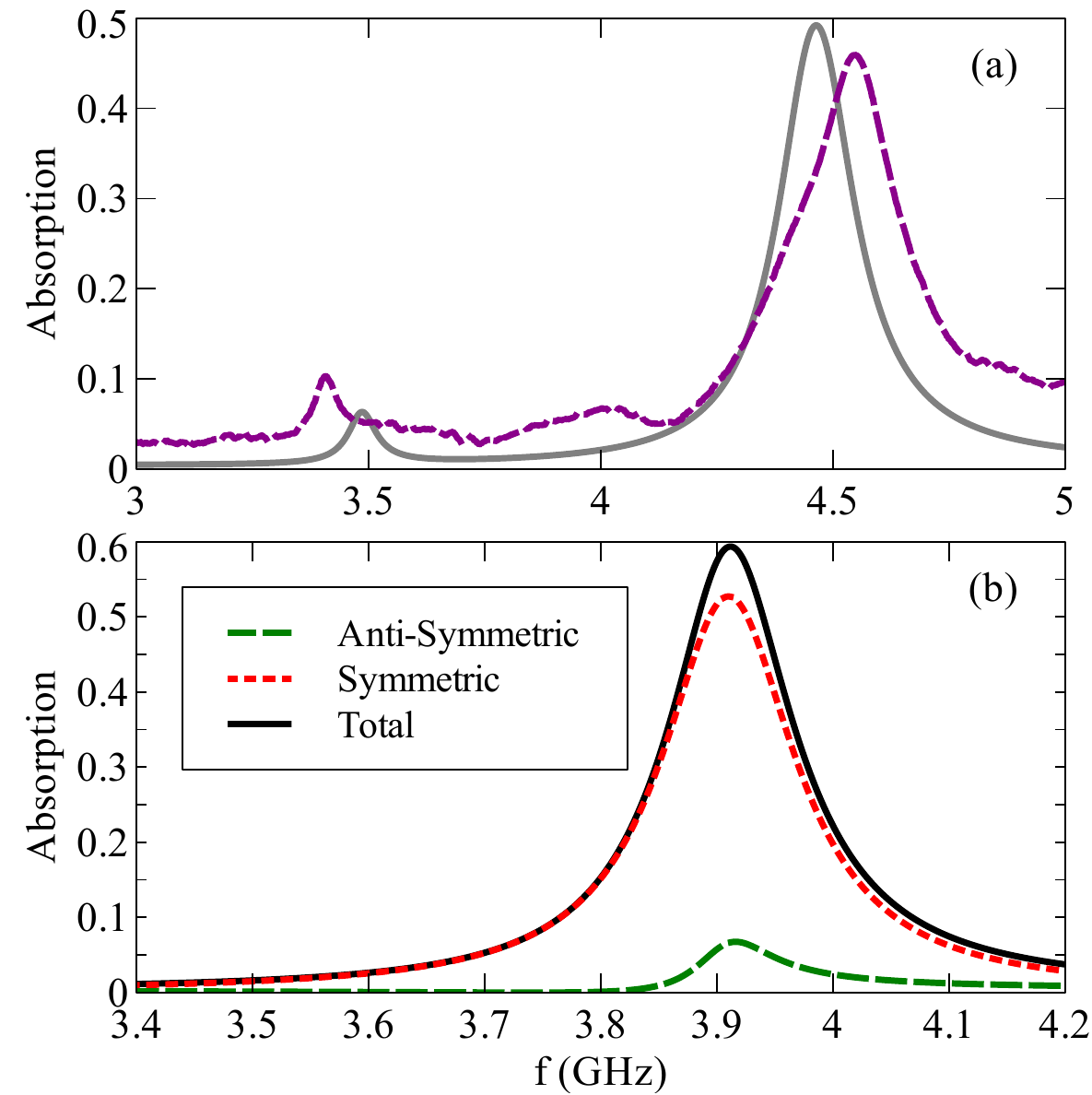}
	\caption{(a) Numerical (solid) and experimental (dashed) absorption curves for $\theta = 180^\circ$. (b) Numerical absorption curve (solid) for $\theta=34^\circ$ at the crossing, showing the two fitted resonances (dashed).}
	\label{fig:absorption_examples}
\end{figure}

For $\theta = 0^\circ$, there are two resonances $\omega_{S}$ and $\omega_{AS}$, and by inspection of the simulated currents in the rings we verify that these correspond to the expected symmetric and antisymmetric modes.  As $\theta$ increases, $\omega_{AS}$  increases and $\omega_{S}$ decreases, reaching their maximum and minimum values respectively at $\theta=180^{\circ}$. As can be seen in Fig.~\ref{fig:experimental_dispersion}(b) and Fig.~\ref{fig:absorption_examples}(a), the two resonant peaks have different widths, which are primarily due to differing radiation losses.  The symmetric mode has relatively stronger radiation losses for low angles, since each ring approximates an electric dipole, and a pair of parallel dipoles radiate strongly.  As the angle approaches 180$^\circ$, the dipoles become oppositely directed, thus we have the low radiation efficiency of an electric quadrupole/magnetic dipole like distribution\cite{Zeng2010}.  The antisymmetric mode has the charges on one ring of the opposite sign to the other, therefore it changes from an electric quadrupole to an electric dipole type distribution with increasing angle, and the radiation efficiency increases.

Inspection of Fig.~\ref{fig:experimental_dispersion}(a) shows that the resonances appear to cross at $\theta\approx 34^{\circ}$.  We can be confident that there is not a very narrow avoided crossing, by considering the corresponding numerical loss coefficients shown in Fig.~\ref{fig:experimental_dispersion}(b), which are clearly distinct from each other near the crossing.  The numerical analysis was repeated in 0.5$^\circ$ steps near this crossing to verify that no features have been missed, thus there is \emph{a clear indication of a crossing of modes}.  In Fig.~\ref{fig:absorption_examples}(b) we have plotted the numerical absorption spectrum in the vicinity of the crossing.  It has the appearance of only a single peak, however the fitting procedure still identifies two separate resonances, which are also plotted for comparison purposes. In general the experimental data show good agreement with the numerical data, but with some uncertainties due to fabrication tolerances.  We will show in Section \ref{sec:morse_theory} how experimental uncertainties contribute to the crossing behavior, but first we will investigate the interaction mechanisms in an ideal system.

\section{Theory of Crossing} \label{sec:theory}
\subsection{Identical and Lossless Rings}\label{sec:ideal_theory}

The tuning of the system by rotation can be explained by looking at the interaction between the rings.  As the rings are twisted, the magnetic and electric near-fields between the two rings change, changing the coupling between them.  We first approach this problem using the Lagrangian for a pair of identical and lossless resonators~\cite{Liu2009a,Powell2010}
\begin{equation} \label{eq:Lagrangian}
\mathcal{L}=A(\dot{Q}_1^2+\dot{Q}_2^2+2\alpha
\dot{Q}_1\dot{Q}_2)-B(Q_1^2+Q_2^2+2\beta Q_1Q_2),
\end{equation}
where $\alpha$ and $\beta$ are the dimensionless magnetic and electric interaction constants, and $Q_{1,2}(t)$ are the time-dependent amplitudes of the modes' charge distributions.  By substituting Eq.~(\ref{eq:Lagrangian}) into the Euler-Lagrange equation the dynamic equations are found to be
\begin{eqnarray}
\ddot{Q}_1+\omega_0^2 Q_1 &=&-\alpha \ddot{Q}_2-\beta \omega_0^2 Q_2.\nonumber\\
\ddot{Q}_2+\omega_0^2 Q_2 &=&-\alpha \ddot{Q}_1-\beta \omega_0^2 Q_1. \label{eq:dynamics_identical}
\end{eqnarray}
Solving the characteristic equation for this system gives two resonances: symmetric ($Q_1 = Q_2$), and antisymmetric ($Q_1 = -Q_2$):
\begin{equation}
\omega_{S} = \omega_{0}\sqrt{\frac{1 + \beta}{1 + \alpha}}, \qquad \omega_{AS} = \omega_{0}\sqrt{\frac{1 - \beta}{1 - \alpha}}.
\label{eq:omega_s_as}
\end{equation}

\begin{figure}[bth]
	\centering
		\includegraphics[width=\columnwidth]{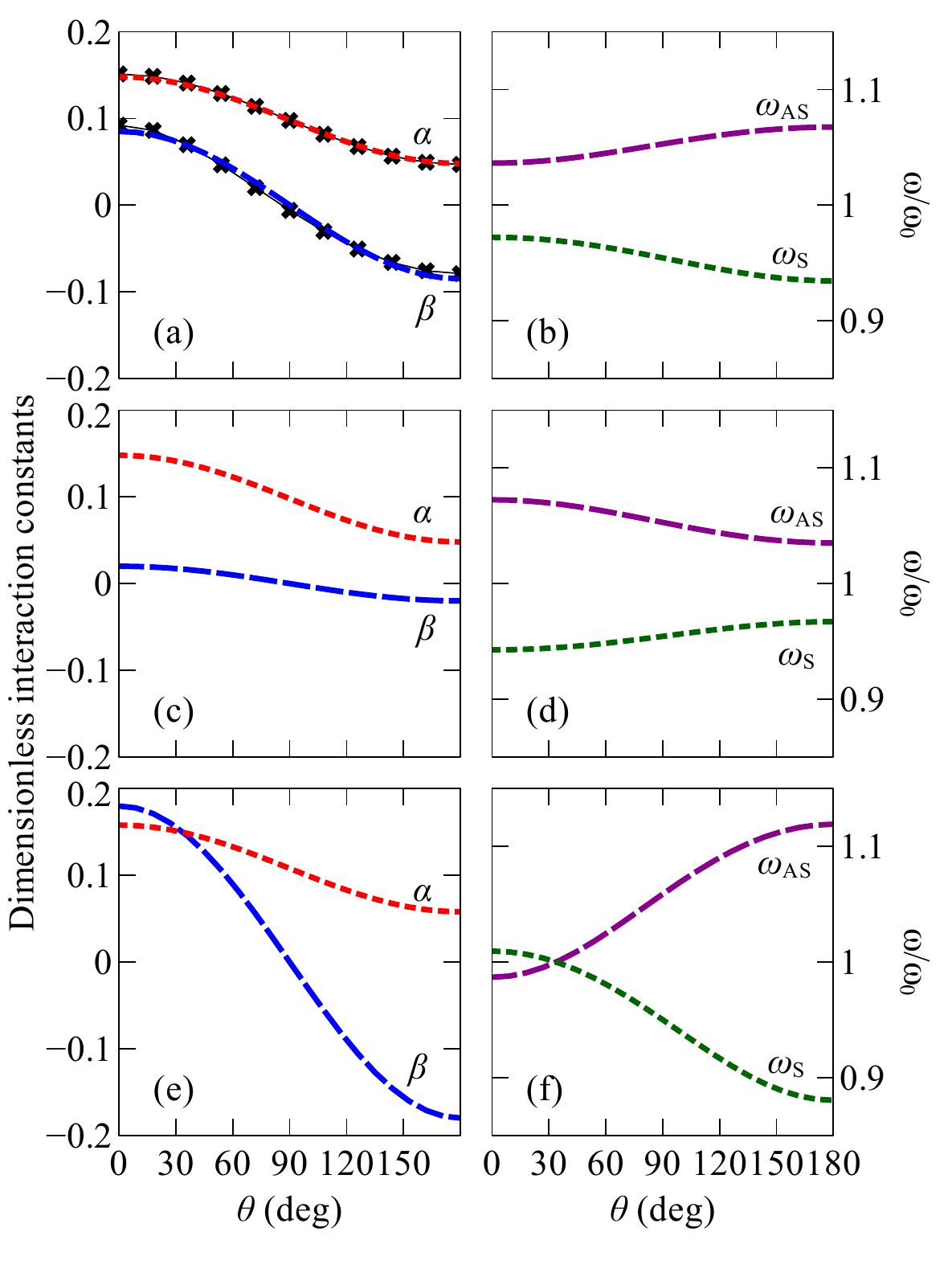}
	\caption{(a) Magnetic ($\alpha$) and electric ($\beta$) interaction constants calculated for a pair of rings in free space.  Dots show exact calculations, lines give the fitted function, and (b) shows the corresponding resonant frequencies. (c) Interaction when electric coupling dominates at $\theta=0^\circ$, and (d) corresponding resonant frequencies.  (e) Interaction constants which become equal, and (f) corresponding resonant frequencies which cross.}
	\label{fig:interaction_all}
\end{figure}

For a pair of rings in \emph{a homogeneous dielectric background}, the electric and magnetic interaction constants can be determined from the interaction energy between resonators, using the method described in Ref.~\onlinecite{Powell2010}.  The resulting interaction constants are shown by the markers in Fig.~\ref{fig:interaction_all}(a), along with the following functions which fit the data very well:
\begin{equation} \label{eq:alpha_beta}
\beta = \beta_{1}\cos(\theta) \qquad \alpha = \alpha_{0} + \alpha_{1}\cos(\theta)
\end{equation}
with $\beta_1 =0.085$, $\alpha_0 =0.098$ and $\alpha_1 =0.05$. These constants are dictated by the charge separation across the gap of the ring, the current circulating around the ring, and the inhomogeneity of the current distribution around the ring, respectively.

For rings aligned on the same axis, we expect that the magnetic interaction $\alpha$ should always be positive, as the intersecting magnetic field from one loop should always be normal to the other loop.  In addition the electric interaction $\beta$ should be positive at $\theta=0^\circ$ as the charge distribution has the nature of parallel dipoles. All arrangement of rings on the same axis which we considered behaved in this manner.

In Fig.~\ref{fig:interaction_all}(b) we plot the corresponding frequencies of the symmetric and antisymmetric modes, normalized to $\omega_0$.  As our approach models the response of the resonators in a homogeneous dielectric background, the results are significantly different from those observed experimentally, where the dielectric is inhomogeneous and the effect of waveguide boundaries is also significant. In particular, for this system of perfectly conducting rings, the crossing of resonances cannot be reproduced for rings in a homogeneous background. Therefore, we consider the possible regimes of interaction which may occur, under the assumption that the interaction constants will be of the form described in Eq.~(\ref{eq:alpha_beta}).

The case considered in Figs.~\ref{fig:interaction_all}(a,b) corresponds to the magnetic interaction always being larger than the electric interaction.  This results in increasing splitting of $\omega_S$ and $\omega_{AS}$ with increasing twist angle, however in principle there is no reason why the splitting cannot decrease.  We show such a case in Figs.~\ref{fig:interaction_all}(c,d), where we have set $\beta_{1}=0.02<\alpha_1$, such that the inhomogeneity in the current has a stronger influence than the dipole-like charge distribution.  Despite the different behavior of the frequency splitting curves, there is little qualitative difference between the interaction constants shown in Figs.~\ref{fig:interaction_all}(a) and (c).

The only other case allowed in our model of interaction under the afore-mentioned physical constraints on $\alpha$ and $\beta$ is that $\alpha > \beta$ for $\theta=0^\circ$.  An example of this is given in Fig.~\ref{fig:interaction_all}(e), where we have set $\alpha_0=0.108, \alpha_1=0.05, \beta_{1}= 0.18$, such that $\beta=\alpha$ at $\theta\approx34^\circ$.  The corresponding resonant frequencies normalized to $\omega_0$ are plotted in Fig.~\ref{fig:interaction_all}(f).    The parameters have been chosen to closely match the dispersion shown in Fig.~\ref{fig:experimental_dispersion}(a).  From this match between the model and the experimental data, we conclude that the inhomogeneous dielectric enhances the electric interaction between the rings but has almost no influence on the magnetic interaction, as expected.

Equations~(\ref{eq:omega_s_as}) show that the tuning curves arise from competition between electric and magnetic interaction constants.  If the magnetic and electric interaction constants have the same sign, then they will counteract each other.  Thus the frequency splitting can be weak even if the near-field interaction is strong.  In particular, the model shows that there is no splitting when $\alpha=\beta$, in agreement with the numerical results presented in Fig.~\ref{fig:experimental_dispersion}.

\subsection{Non-identical and Lossy Rings} \label{sec:morse_theory}

Up to this point, our analysis has been based on the assumption that the two rings are absolutely identical, which means that the hybridized modes will be symmetric and antisymmetric, and degeneracy can occur when the effective coupling is canceled.  If a pair of lossless rings are not absolutely identical then their hybridized modes will repel each other and the crossing will be avoided.  In our system we expect that the fabricated rings would never be perfectly identical, and their different orientation with respect to the waveguide would give them slightly different resonant frequencies, thus preventing degeneracy.  However, our system also has significant losses, primarily due to radiation into the forward and backward propagating waveguide modes, but also some absorption due to metallic and dielectric losses.

To analyze the more realistic model which includes detuning and losses, it is convenient to use the theory of Morse critical points, which has previously been applied to resonators and waveguides\cite{Shestopalov1987,Shestopalov1993,Yakovlev1997}.  In Ref.~\onlinecite{Yakovlev2000} an approach was demonstrated which showed the conditions for crossing or anti-crossing of modes, and in particular demonstrated that losses can counteract an avoided crossing. We apply this approach to our system and present the key results.  For a complete derivation and theoretical background, the reader is referred to Ref.~\onlinecite{Yakovlev2000}.
First we modify the system of equations given by Eqs.~(\ref{eq:dynamics_identical}) by introducing a dissipation coefficient $\Gamma$ and a detuning parameter $\delta\omega$:
\begin{eqnarray}
\ddot{Q}_1+2\Gamma\dot{Q}_1+\left(\omega_0+\delta\omega\right)^2 Q_1 &=&-\alpha \ddot{Q}_2-\beta \omega_0^2 Q_2.\nonumber\\
\ddot{Q}_2+2\Gamma\dot{Q}_2+\left(\omega_0-\delta\omega\right)^2 Q_2 &=&-\alpha \ddot{Q}_1-\beta \omega_0^2 Q_1. \label{eq:dynamics_lossy}
\end{eqnarray}
By taking the time dependence of $Q_{1,2}$ as $\exp(j\omega t)$, we arrive at the following dispersion equation:
\begin{eqnarray}
D(\omega, \theta) =
\left[\left(\omega_0+\delta\omega\right)^2 + j2\Gamma \omega - \omega^2\right] \nonumber\\
\times \left[\left(\omega_0-\delta\omega\right)^2 + j2\Gamma \omega - \omega^2\right] \nonumber\\
-\left[\beta(\theta) \omega_0^2 - \alpha(\theta) \omega^2\right]^2 = 0
\end{eqnarray}
The solutions of this system are not strictly symmetric and antisymmetric, although for small $\delta\omega$ and for angles away from the avoided-crossing they are only slightly perturbed from the original modes. This model neglects the differences in radiation losses of the symmetric and antisymmetric modes, which also vary with twist angle. For this example we take a detuning parameter $\delta\omega = 10^{-3}\omega_{0}$, and $\alpha$ and $\beta$ are given by Eq.~(\ref{eq:alpha_beta}), with the same coefficients used to derive Figs.~\ref{fig:interaction_all}(e,f).

The dispersion curves of the system correspond to the condition $D(\omega, \theta) = 0$. In applying the theory of Morse critical points, we study the behavior of the function $D(\omega, \theta)$ in the neighborhood of the crossing/anti-crossing, and do not limit ourselves to values of $(\omega, \theta)$ which satisfy the dispersion equation.

The first step is to find the Morse critical point $(\omega_{m}, \theta_{m})$ which satisfies $D'_{\omega}(\omega_{m}, \theta_{m}) = D'_{\theta}(\omega_{m}, \theta_{m}) = 0$.  In the case of a crossing of modes, this will be the point where the dispersion curves cross, in the case of an avoided crossing this will be a saddle-point of $D(\omega, \theta)$. In general this point must be found numerically, however for our system of equations we find a sufficiently accurate solution by using perturbation theory to first order in $\Gamma$:
\begin{eqnarray}
\omega_{m} &=& \omega_{m0} + j\Gamma \label{eq:omega_m}\\
\omega_{m0} &=& \left(\omega_0^2+\delta\omega^2\right)^{\frac{1}{2}} \\
\cos(\theta_{m}) &=& \frac{\alpha_{0}\omega_{m0}^2}{\alpha_{1}\omega_{m0}^2-\beta_{1}\omega_{0}^{2}} +
 j\Gamma\frac{2\alpha_0\beta_1\omega_{m0}\omega_0^2}{\left(\alpha_{1}\omega_{m0}^2-\beta_{1}\omega_0^2\right)^2} \label{eq:theta_m}
\end{eqnarray}

In the vicinity of the Morse point we perform a second-order Taylor expansion of $D(\omega, \theta)$
\begin{eqnarray}
D(\omega, \theta) = D(\omega_{m}, \theta_{m}) + D_{\omega\omega}''(\omega-\omega_{m})^2/2 + \nonumber\\
D_{\omega\theta}''(\omega-\omega_{m})(\theta-\theta_{m}) + D_{\theta\theta}''(\theta-\theta_{m})^2/2.
\end{eqnarray}
For our system the relevant partial derivatives can be calculated analytically, but the expressions are long and not particularly illuminating.
From this local form of the dispersion equation, the solutions are given by:
\begin{eqnarray}
\omega(\theta) = \omega_m - \frac{D''_{\omega\theta}}{D_{\omega\omega}''}(\theta-\theta_{m}) \pm \left(D_{\omega\omega}''\right)^{-1}\times\nonumber\\
\left[\left({D''_{\omega\theta}}^2\!-\! D_{\omega\omega}''D_{\theta\theta}''\right)\left(\theta-\theta_m\right)^2\!-\!2D_{\theta\theta}''D(\omega_{m}, \theta_{m})\right]^{\frac{1}{2}}
\end{eqnarray}
This function has a pair of branch points, where the argument of the square root becomes zero. Upon substitution of the parameters of our system, we find that the branch points occur at the following values\footnote{Strictly speaking the term inside the square root has some dependence on $\Gamma$, which in turn will cause it to become complex for $\Gamma\neq0$.  However both effects are negligible in our system.} of $\theta$:
\begin{equation} \label{eq:branch_points}
\theta_{b1,2} = \theta_m \pm 2j\delta\omega\omega_{0}\frac{\sqrt{
\left(\alpha_0\beta_1\omega_0^2\right)^2-\left(\alpha_1^2\omega_{m0}^2-\beta_1\omega_0^2\right)^2
}}
{\left(\alpha_1\omega_{m0}^2-\beta_1\omega_0^2\right)^2}.
\end{equation}

As shown in Ref.~\onlinecite{Yakovlev2000}, the position of branch points $\theta_{b1,2}$ on the complex plane determines the behavior of the resonant frequencies of the modes. In brief, if the imaginary parts of the two branch points have different signs, then the modes undergo an anti-crossing. However, the modes cross if the imaginary parts of the $\theta_{b1,2}$ have the same sign. This allows us to analyze the effect of detuning $\delta \omega$ and losses $\Gamma$ on the crossing--anti-crossing behavior of the system. To do this, we fix the value of detuning $\delta\omega=10^{-3}\omega_0$ and change the losses $\Gamma$.   
   
\begin{figure}[bt]
	\centering
	\includegraphics[width=\columnwidth]{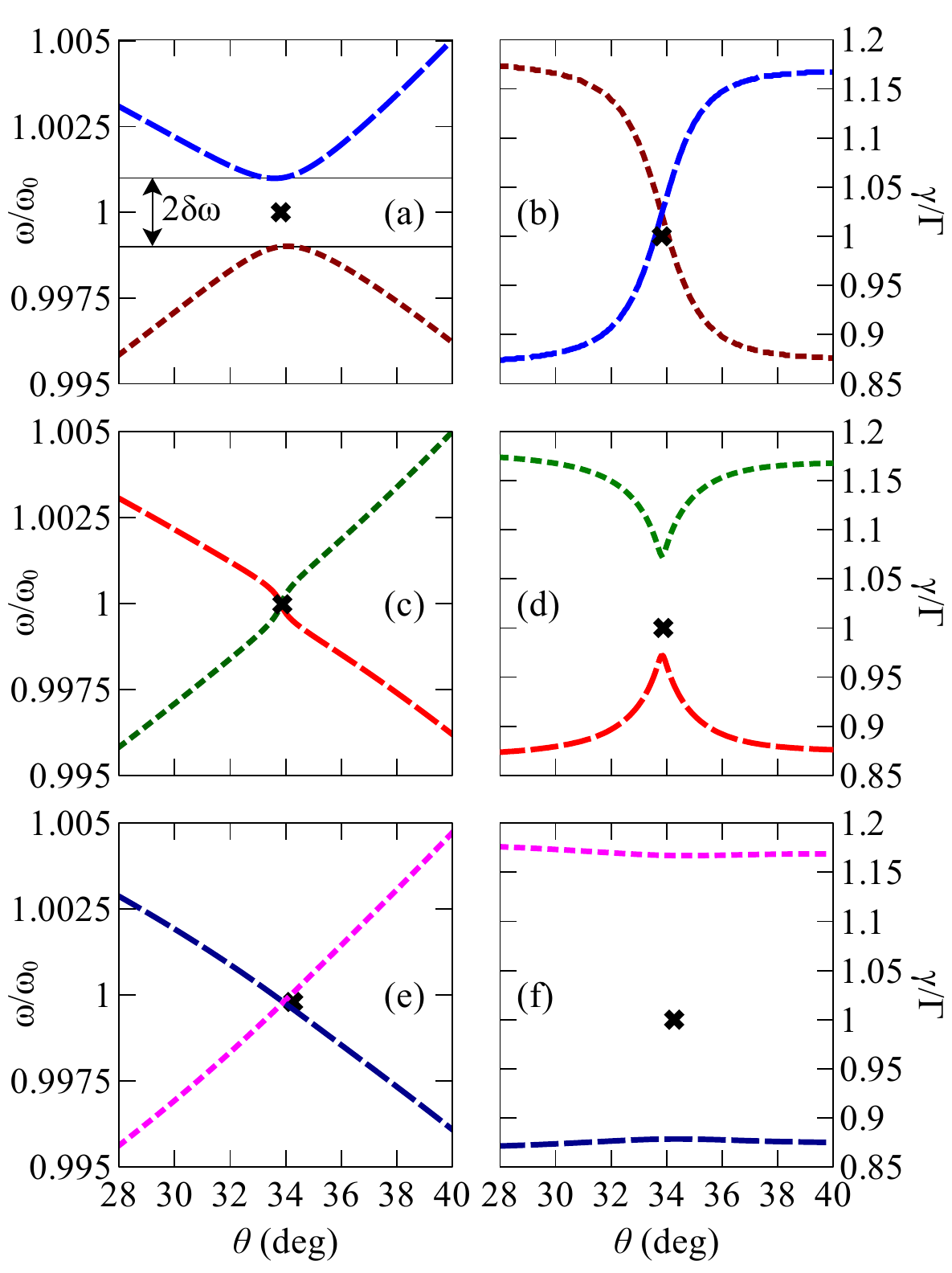}
	\caption{(a) Resonant frequencies and (b) absorption coefficients when $\Gamma=1\times10^{-3}\omega_{0}$, (c) Resonant frequencies and (d) absorption coefficients when $\Gamma=7\times10^{-3}\omega_{0}$, showing that losses restore the crossing.  (e) Resonant frequencies and (f) absorption coefficients when losses are increased to $\Gamma=2\times10^{-2}\omega_{0}$.  The projections of the Morse point are shown by crosses.}
	\label{fig:crossing_analytical}
\end{figure}

We start with weak losses, when $\Gamma=1\times10^{-3}\omega_{0}$.  This case is very close to lossless regime, and the real parts of frequencies are plotted in Fig.~\ref{fig:crossing_analytical}(a), where we see an avoided crossing, and in Fig.~\ref{fig:crossing_analytical}(b) we see the imaginary parts, which are exchanged.

On the other hand, if we increase losses sufficiently, we observe a fundamental change in the behavior of the dispersion curves, and the crossing of the real parts of the eigenmodes appears. In our system we set $\Gamma=7\times10^{-3}\omega_{0}$, and the resulting dispersion and absorption curves can be seen in Fig.~\ref{fig:crossing_analytical}(c) and (d) respectively. If we increase $\Gamma$ further to $1\times10^{-2}\omega_{0}$, the curves become smoother in the vicinity of the Morse critical point, showing similar character to the numerical data [see Fig.~\ref{fig:crossing_analytical}(e) and (f)]. Since to first order $\omega_i=\Gamma$, this case is comparable to the experimental and numerical results in Fig.~\ref{fig:experimental_dispersion}. 

We have shown how degeneracy (mode crossing) can arise in the experimental data, even when the rings are not absolutely identical.  According to our model, the threshold of losses which will mitigate the avoided crossing ($\Gamma/\omega_{0}\approx7\times10^{-3}$) is similar to the experimentally measured values.  However this is based on the assumption that the resonant frequencies of the two rings varied by no more than 0.1\% ($\delta_\omega=10^{-3}\omega_0$).  The fabrication and alignment tolerances in our system are greater than this, thus we infer that there is a small avoided crossing in our system.  It is not possible to measure this directly from experiment, since each angle corresponds to a different sample with different fabrication errors.

From the numerical results in Fig.~\ref{fig:experimental_dispersion}(c) and Fig.~\ref{fig:crossing_analytical}(d), we see that even without any fabrication error, the decay constants of the hybridized modes are quite different, due to different radiation losses.  We note that if complete degeneracy of both the frequency and decay constants of the eigenmodes could be achieved, then the coupling between the rings would be completely suppressed.  This would have important implications for mitigating spatial-dispersion, which often inhibits the ability of bulk metamaterials to achieve specified properties.

\section{Conclusion} \label{sec:conclusion}

We have shown in a series of experiments how changing the relative rotation between two rings modifies the electric and magnetic interaction between them, thus tuning the hybridized resonances.  Using both numerical simulations and an analytical model, that takes into account both electric and magnetic interactions between the rings, we have shown that there is a crossing where the two resonances coexist, at an angle where the electric and magnetic coupling are equal.  We have shown how experimental errors can cause the crossing of the dispersion curve to be avoided, however using the theory of Morse critical points, we have demonstrated that losses can cause the crossing to be restored.

\begin{acknowledgments}
We acknowledge funding from the Australian Research Council.
\end{acknowledgments}


\begin{thebibliography}{10}%
\makeatletter
\providecommand \@ifxundefined [1]{%
 \ifx #1\undefined \expandafter \@firstoftwo
 \else \expandafter \@secondoftwo
\fi
}%
\providecommand \@ifnum [1]{%
 \ifnum #1\expandafter \@firstoftwo
 \else \expandafter \@secondoftwo
\fi
}%
\providecommand \enquote [1]{``#1''}%
\providecommand \bibnamefont  [1]{#1}%
\providecommand \bibfnamefont [1]{#1}%
\providecommand \citenamefont [1]{#1}%
\providecommand\href[0]{\@sanitize\@href}%
\providecommand\@href[1]{\endgroup\@@startlink{#1}\endgroup\@@href}%
\providecommand\@@href[1]{#1\@@endlink}%
\providecommand \@sanitize [0]{\begingroup\catcode`\&12\catcode`\#12\relax}%
\@ifxundefined \pdfoutput {\@firstoftwo}{%
 \@ifnum{\z@=\pdfoutput}{\@firstoftwo}{\@secondoftwo}%
}{%
 \providecommand\@@startlink[1]{\leavevmode}%
 \providecommand\@@endlink[0]{}%
}{%
 \providecommand\@@startlink[1]{%
  \leavevmode
  \pdfstartlink
   attr{/Border[0 0 1 ]/H/I/C[0 1 1]}%
   user{/Subtype/Link/A<</Type/Action/S/URI/URI(#1)>>}%
  \relax
 }%
 \providecommand\@@endlink[0]{\pdfendlink}%
}%
\providecommand \url  [0]{\begingroup\@sanitize \@url }%
\providecommand \@url [1]{\endgroup\@href {#1}{\urlprefix}}%
\providecommand \urlprefix [0]{URL }%
\providecommand \Eprint[0]{\href }%
\@ifxundefined \urlstyle {%
  \providecommand \doi [1]{doi:\discretionary{}{}{}#1}%
}{%
  \providecommand \doi [0]{doi:\discretionary{}{}{}\begingroup
  \urlstyle{rm}\Url }%
}%
\providecommand \doibase [0]{http://dx.doi.org/}%
\providecommand \Doi[1]{\href{\doibase#1}}%
\providecommand \bibAnnote [3]{%
  \BibitemShut{#1}%
  \begin{quotation}\noindent
    \textsc{Key:}\ #2\\\textsc{Annotation:}\ #3%
  \end{quotation}%
}%
\providecommand \bibAnnoteFile [2]{%
  \IfFileExists{#2}{\bibAnnote {#1} {#2} {\input{#2}}}{}%
}%
\providecommand \typeout [0]{\immediate \write \m@ne }%
\providecommand \selectlanguage [0]{\@gobble}%
\providecommand \bibinfo [0]{\@secondoftwo}%
\providecommand \bibfield [0]{\@secondoftwo}%
\providecommand \translation [1]{[#1]}%
\providecommand \BibitemOpen[0]{}%
\providecommand \bibitemStop [0]{}%
\providecommand \bibitemNoStop [0]{.\EOS\space}%
\providecommand \EOS [0]{\spacefactor3000\relax}%
\providecommand \BibitemShut [1]{\csname bibitem#1\endcsname}%
\bibitem{Smith2000}%
  \BibitemOpen
  \bibfield{author}{%
  \bibinfo {author} {\bibfnamefont{D.}~\bibnamefont{Smith}}, \bibinfo {author}
  {\bibfnamefont{W.~J.}\ \bibnamefont{Padilla}}, \bibinfo {author}
  {\bibfnamefont{D.}~\bibnamefont{Vier}}, \bibinfo {author}
  {\bibfnamefont{S.}~\bibnamefont{Nemat-Nasser}},\ and\ \bibinfo {author}
  {\bibfnamefont{S.}~\bibnamefont{Schultz}},\ }%
  \bibfield{journal}{%
  \bibinfo {journal} {{Phys. Rev. Lett.}}\ }%
  \textbf{\bibinfo {volume} {{84}}},\ \bibinfo {pages} {{4184}} (\bibinfo
  {year} {{2000}})%
  \bibAnnoteFile{NoStop}{Smith2000}%
\bibitem{Pendry1999}%
  \BibitemOpen
  \bibfield{author}{%
  \bibinfo {author} {\bibfnamefont{J.}~\bibnamefont{Pendry}}, \bibinfo {author}
  {\bibfnamefont{A.}~\bibnamefont{Holden}}, \bibinfo {author}
  {\bibfnamefont{D.}~\bibnamefont{Robbins}},\ and\ \bibinfo {author}
  {\bibfnamefont{W.}~\bibnamefont{Stewart}},\ }%
  \bibfield{journal}{%
  \bibinfo {journal} {{IEEE Trans. Microwave Theory Tech.}}\ }%
  \textbf{\bibinfo {volume} {{47}}},\ \bibinfo {pages} {{2075}} (\bibinfo
  {year} {{1999}})%
  \bibAnnoteFile{NoStop}{Pendry1999}%
\bibitem{Gay-Balmaz2002}%
  \BibitemOpen
  \bibfield{author}{%
  \bibinfo {author} {\bibfnamefont{P.}~\bibnamefont{Gay-Balmaz}}\ and\ \bibinfo
  {author} {\bibfnamefont{O.~J.~F.}\ \bibnamefont{Martin}},\ }%
  \bibfield{journal}{%
  \Doi{10.1063/1.1497452}{\bibinfo {journal} {J. Appl. Phys.}}\ }%
  \textbf{\bibinfo {volume} {92}},\ \bibinfo {pages} {2929} (\bibinfo {year}
  {2002})%
  \bibAnnoteFile{NoStop}{Gay-Balmaz2002}%
\bibitem{Gorkunov2002}%
  \BibitemOpen
  \bibfield{author}{%
  \bibinfo {author} {\bibfnamefont{M.}~\bibnamefont{Gorkunov}}, \bibinfo
  {author} {\bibfnamefont{M.}~\bibnamefont{Lapine}}, \bibinfo {author}
  {\bibfnamefont{E.}~\bibnamefont{Shamonina}},\ and\ \bibinfo {author}
  {\bibfnamefont{K.}~\bibnamefont{Ringhofer}},\ }%
  \bibfield{journal}{%
  \Doi{10.1140/epjb/e2002-00228-4}{\bibinfo {journal} {Eur. Phys. J. B}}\ }%
  \textbf{\bibinfo {volume} {28}},\ \bibinfo {pages} {263} (\bibinfo {year}
  {2002})%
  \bibAnnoteFile{NoStop}{Gorkunov2002}%
\bibitem{Lapine2009}%
  \BibitemOpen
  \bibfield{author}{%
  \bibinfo {author} {\bibfnamefont{M.}~\bibnamefont{Lapine}}, \bibinfo {author}
  {\bibfnamefont{D.}~\bibnamefont{Powell}}, \bibinfo {author}
  {\bibfnamefont{M.}~\bibnamefont{Gorkunov}}, \bibinfo {author}
  {\bibfnamefont{I.}~\bibnamefont{Shadrivov}}, \bibinfo {author}
  {\bibfnamefont{R.}~\bibnamefont{Marques}},\ and\ \bibinfo {author}
  {\bibfnamefont{Y.}~\bibnamefont{Kivshar}},\ }%
  \bibfield{journal}{%
  \Doi{{10.1063/1.3211920}}{\bibinfo {journal} {{App. Phys. Lett.}}}\ }%
  \textbf{\bibinfo {volume} {{95}}},\ \bibinfo {pages} {{084105}} (\bibinfo
  {year} {{2009}})%
  \bibAnnoteFile{NoStop}{Lapine2009}%
\bibitem{Powell2010}%
  \BibitemOpen
  \bibfield{author}{%
  \bibinfo {author} {\bibfnamefont{D.~A.}\ \bibnamefont{Powell}}, \bibinfo
  {author} {\bibfnamefont{M.}~\bibnamefont{Lapine}}, \bibinfo {author}
  {\bibfnamefont{M.}~\bibnamefont{Gorkunov}}, \bibinfo {author}
  {\bibfnamefont{I.~V.}\ \bibnamefont{Shadrivov}},\ and\ \bibinfo {author}
  {\bibfnamefont{Y.~S.}\ \bibnamefont{Kivshar}},\ }%
  \bibfield{journal}{%
  \Doi{10.1103/PhysRevB.82.155128}{\bibinfo {journal} {Phys. Rev. B}}\ }%
  \textbf{\bibinfo {volume} {82}},\ \bibinfo {pages} {155128} (\bibinfo {year}
  {2010})%
  \bibAnnoteFile{NoStop}{Powell2010}%
\bibitem{Feth2010}%
  \BibitemOpen
  \bibfield{author}{%
  \bibinfo {author} {\bibfnamefont{N.}~\bibnamefont{Feth}}, \bibinfo {author}
  {\bibfnamefont{M.}~\bibnamefont{K\"{o}nig}}, \bibinfo {author}
  {\bibfnamefont{M.}~\bibnamefont{Husnik}}, \bibinfo {author}
  {\bibfnamefont{K.}~\bibnamefont{Stannigel}}, \bibinfo {author}
  {\bibfnamefont{J.}~\bibnamefont{Niegemann}}, \bibinfo {author}
  {\bibfnamefont{K.}~\bibnamefont{Busch}}, \bibinfo {author}
  {\bibfnamefont{M.}~\bibnamefont{Wegener}},\ and\ \bibinfo {author}
  {\bibfnamefont{S.}~\bibnamefont{Linden}},\ }%
  \bibfield{journal}{%
  \bibinfo {journal} {Opt. Express}\ }%
  \textbf{\bibinfo {volume} {18}},\ \bibinfo {pages} {6545} (\bibinfo {year}
  {2010})%
  \bibAnnoteFile{NoStop}{Feth2010}%
\bibitem{Sersic2009}%
  \BibitemOpen
  \bibfield{author}{%
  \bibinfo {author} {\bibfnamefont{I.}~\bibnamefont{Sersic}}, \bibinfo {author}
  {\bibfnamefont{M.}~\bibnamefont{Frimmer}}, \bibinfo {author}
  {\bibfnamefont{E.}~\bibnamefont{Verhagen}},\ and\ \bibinfo {author}
  {\bibfnamefont{A.~F.}\ \bibnamefont{Koenderink}},\ }%
  \bibfield{journal}{%
  \Doi{10.1103/PhysRevLett.103.213902}{\bibinfo {journal} {Phys. Rev. Lett.}}\
  }%
  \textbf{\bibinfo {volume} {103}},\ \bibinfo {pages} {213902} (\bibinfo {year}
  {2009})%
  \bibAnnoteFile{NoStop}{Sersic2009}%
\bibitem{Liu2008}%
  \BibitemOpen
  \bibfield{author}{%
  \bibinfo {author} {\bibfnamefont{N.}~\bibnamefont{Liu}}\ and\ \bibinfo
  {author} {\bibfnamefont{H.}~\bibnamefont{Giessen}},\ }%
  \bibfield{journal}{%
  \bibinfo {journal} {Opt. Express}\ }%
  \textbf{\bibinfo {volume} {16}},\ \bibinfo {pages} {21233} (\bibinfo {year}
  {2008})%
  \bibAnnoteFile{NoStop}{Liu2008}%
\bibitem{Liu2008a}%
  \BibitemOpen
  \bibfield{author}{%
  \bibinfo {author} {\bibfnamefont{N.}~\bibnamefont{Liu}}, \bibinfo {author}
  {\bibfnamefont{S.}~\bibnamefont{Kaiser}},\ and\ \bibinfo {author}
  {\bibfnamefont{H.}~\bibnamefont{Giessen}},\ }%
  \bibfield{journal}{%
  \Doi{10.1002/adma.200801917}{\bibinfo {journal} {Adv. Mater.}}\ }%
  \textbf{\bibinfo {volume} {20}},\ \bibinfo {pages} {4521} (\bibinfo {year}
  {2008})%
  \bibAnnoteFile{NoStop}{Liu2008a}%
\bibitem{Liu2009a}%
  \BibitemOpen
  \bibfield{author}{%
  \bibinfo {author} {\bibfnamefont{H.}~\bibnamefont{Liu}}, \bibinfo {author}
  {\bibfnamefont{Y.~M.}\ \bibnamefont{Liu}}, \bibinfo {author}
  {\bibfnamefont{T.}~\bibnamefont{Li}}, \bibinfo {author}
  {\bibfnamefont{S.~M.}\ \bibnamefont{Wang}}, \bibinfo {author}
  {\bibfnamefont{S.~N.}\ \bibnamefont{Zhu}},\ and\ \bibinfo {author}
  {\bibfnamefont{X.}~\bibnamefont{Zhang}},\ }%
  \bibfield{journal}{%
  \Doi{10.1002/pssb.200844414}{\bibinfo {journal} {Phys. Status Solidi B}}\ }%
  \textbf{\bibinfo {volume} {246}},\ \bibinfo {pages} {1397} (\bibinfo {year}
  {2009})%
  \bibAnnoteFile{NoStop}{Liu2009a}%
\bibitem{Liu2007}%
  \BibitemOpen
  \bibfield{author}{%
  \bibinfo {author} {\bibfnamefont{H.}~\bibnamefont{Liu}}, \bibinfo {author}
  {\bibfnamefont{D.}~\bibnamefont{Genov}}, \bibinfo {author}
  {\bibfnamefont{D.}~\bibnamefont{Wu}}, \bibinfo {author}
  {\bibfnamefont{Y.}~\bibnamefont{Liu}}, \bibinfo {author}
  {\bibfnamefont{Z.}~\bibnamefont{Liu}}, \bibinfo {author}
  {\bibfnamefont{C.}~\bibnamefont{Sun}}, \bibinfo {author}
  {\bibfnamefont{S.}~\bibnamefont{Zhu}},\ and\ \bibinfo {author}
  {\bibfnamefont{X.}~\bibnamefont{Zhang}},\ }%
  \bibfield{journal}{%
  \Doi{{10.1103/PhysRevB.76.073101}}{\bibinfo {journal} {{Phys. Rev. B}}}\ }%
  \textbf{\bibinfo {volume} {{76}}},\ \bibinfo {pages} {{073101}} (\bibinfo
  {year} {{2007}})%
  \bibAnnoteFile{NoStop}{Liu2007}%
\bibitem{Li2009a}%
  \BibitemOpen
  \bibfield{author}{%
  \bibinfo {author} {\bibfnamefont{T.~Q.}\ \bibnamefont{Li}}, \bibinfo {author}
  {\bibfnamefont{H.}~\bibnamefont{Liu}}, \bibinfo {author}
  {\bibfnamefont{T.}~\bibnamefont{Li}}, \bibinfo {author}
  {\bibfnamefont{S.~M.}\ \bibnamefont{Wang}}, \bibinfo {author}
  {\bibfnamefont{J.~X.}\ \bibnamefont{Cao}}, \bibinfo {author}
  {\bibfnamefont{Z.~H.}\ \bibnamefont{Zhu}}, \bibinfo {author}
  {\bibfnamefont{Z.~G.}\ \bibnamefont{Dong}}, \bibinfo {author}
  {\bibfnamefont{S.~N.}\ \bibnamefont{Zhu}},\ and\ \bibinfo {author}
  {\bibfnamefont{X.}~\bibnamefont{Zhang}},\ }%
  \bibfield{journal}{%
  \Doi{10.1103/PhysRevB.80.115113}{\bibinfo {journal} {Phys. Rev. B}}\ }%
  \textbf{\bibinfo {volume} {80}},\ \bibinfo {pages} {115113} (\bibinfo {year}
  {2009})%
  \bibAnnoteFile{NoStop}{Li2009a}%
\bibitem{Liu2009}%
  \BibitemOpen
  \bibfield{author}{%
  \bibinfo {author} {\bibfnamefont{N.}~\bibnamefont{Liu}}, \bibinfo {author}
  {\bibfnamefont{H.}~\bibnamefont{Liu}}, \bibinfo {author}
  {\bibfnamefont{S.}~\bibnamefont{Zhu}},\ and\ \bibinfo {author}
  {\bibfnamefont{H.}~\bibnamefont{Giessen}},\ }%
  \bibfield{journal}{%
  \Doi{{10.1038/nphoton.2009.4}}{\bibinfo {journal} {{Nat. Photon}}}\ }%
  \textbf{\bibinfo {volume} {{3}}},\ \bibinfo {pages} {{157}} (\bibinfo {year}
  {{2009}})%
  \bibAnnoteFile{NoStop}{Liu2009}%
\bibitem{Liu2010}%
  \BibitemOpen
  \bibfield{author}{%
  \bibinfo {author} {\bibfnamefont{H.}~\bibnamefont{Liu}}, \bibinfo {author}
  {\bibfnamefont{J.~X.}\ \bibnamefont{Cao}}, \bibinfo {author}
  {\bibfnamefont{S.~N.}\ \bibnamefont{Zhu}}, \bibinfo {author}
  {\bibfnamefont{N.}~\bibnamefont{Liu}}, \bibinfo {author}
  {\bibfnamefont{R.}~\bibnamefont{Ameling}},\ and\ \bibinfo {author}
  {\bibfnamefont{H.}~\bibnamefont{Giessen}},\ }%
  \bibfield{journal}{%
  \Doi{10.1103/PhysRevB.81.241403}{\bibinfo {journal} {Phys. Rev. B}}\ }%
  \textbf{\bibinfo {volume} {81}},\ \bibinfo {pages} {241403} (\bibinfo {year}
  {2010})%
  \bibAnnoteFile{NoStop}{Liu2010}%
\bibitem{Miroshnichenko2010}%
  \BibitemOpen
  \bibfield{author}{%
  \bibinfo {author} {\bibfnamefont{A.}~\bibnamefont{Miroshnichenko}}, \bibinfo
  {author} {\bibfnamefont{S.}~\bibnamefont{Flach}},\ and\ \bibinfo {author}
  {\bibfnamefont{Y.~S.}\ \bibnamefont{Kivshar}},\ }%
  \bibfield{journal}{%
  \Doi{10.1103/RevModPhys.82.2257}{\bibinfo {journal} {Rev. Mod. Phys.}}\ }%
  \textbf{\bibinfo {volume} {82}},\ \bibinfo {pages} {2257} (\bibinfo {year}
  {2010})%
  \bibAnnoteFile{NoStop}{Miroshnichenko2010}%
\bibitem{Zeng2010}%
  \BibitemOpen
  \bibfield{author}{%
  \bibinfo {author} {\bibfnamefont{Y.}~\bibnamefont{Zeng}}, \bibinfo {author}
  {\bibfnamefont{C.}~\bibnamefont{Dineen}},\ and\ \bibinfo {author}
  {\bibfnamefont{J.~V.}\ \bibnamefont{Moloney}},\ }%
  \bibfield{journal}{%
  \Doi{10.1103/PhysRevB.81.075116}{\bibinfo {journal} {Phys. Rev. B}}\ }%
  \textbf{\bibinfo {volume} {81}},\ \bibinfo {pages} {075116} (\bibinfo {year}
  {2010})%
  \bibAnnoteFile{NoStop}{Zeng2010}%
\bibitem{Shestopalov1987}%
  \BibitemOpen
  \bibfield{author}{%
  \bibinfo {author} {\bibfnamefont{V.~P.}\ \bibnamefont{Shestopalov}},\ }%
  \emph{\bibinfo {title} {Spectral theory and excitation of open structures}}\
  (\bibinfo {publisher} {Naukova Dumka},\ \bibinfo {address} {Kiev},\ \bibinfo
  {year} {1987})\ \bibinfo {note} {in Russian}%
  \bibAnnoteFile{NoStop}{Shestopalov1987}%
\bibitem{Shestopalov1993}%
  \BibitemOpen
  \bibfield{author}{%
  \bibinfo {author} {\bibfnamefont{V.~P.}\ \bibnamefont{Shestopalov}},\ }%
  \bibfield{journal}{%
  \Doi{10.1080/02726349308908348}{\bibinfo {journal} {Electromagnetics}}\ }%
  \textbf{\bibinfo {volume} {13}},\ \bibinfo {pages} {239} (\bibinfo {year}
  {1993})%
  \bibAnnoteFile{NoStop}{Shestopalov1993}%
\bibitem{Yakovlev1997}%
  \BibitemOpen
  \bibfield{author}{%
  \bibinfo {author} {\bibfnamefont{A.~B.}\ \bibnamefont{Yakovlev}}\ and\
  \bibinfo {author} {\bibfnamefont{G.}~\bibnamefont{Hanson}},\ }%
  \bibfield{journal}{%
  \Doi{10.1109/22.552036}{\bibinfo {journal} {IEEE Trans. Microwave Theory
  Tech.}}\ }%
  \textbf{\bibinfo {volume} {45}},\ \bibinfo {pages} {87} (\bibinfo {year}
  {1997})%
  \bibAnnoteFile{NoStop}{Yakovlev1997}%
\bibitem{Yakovlev2000}%
  \BibitemOpen
  \bibfield{author}{%
  \bibinfo {author} {\bibfnamefont{A.~B.}\ \bibnamefont{Yakovlev}}\ and\
  \bibinfo {author} {\bibfnamefont{G.}~\bibnamefont{Hanson}},\ }%
  \bibfield{journal}{%
  \Doi{10.1109/22.817473}{\bibinfo {journal} {IEEE Trans. Microwave Theory
  Tech.}}\ }%
  \textbf{\bibinfo {volume} {48}},\ \bibinfo {pages} {67} (\bibinfo {year}
  {2000})%
  \bibAnnoteFile{NoStop}{Yakovlev2000}%
\bibitem{Note1}%
  \BibitemOpen
  \bibinfo {note} {Strictly speaking the term inside the square root has some
  dependence on $\Gamma $, which in turn will cause it to become complex for
  $\Gamma \not =0$. However both effects are negligible in our system.}%
  \bibAnnoteFile{Stop}{Note1}%
\end{thebibliography}

%

\end{document}